# Lattice mismatch as the descriptor of segregation, stability and reactivity of supported thin catalyst films

Edvin Fako[1,2], Ana S. Dobrota[1], Igor A. Pašti[1] *, Núria López[2], Slavko V. Mentus[1,3], Natalia V. Skorodumova[4,5]

[1]*University of Belgrade – Faculty of Physical Chemistry, Studentski trg 12-16, 11158 Belgrade, Serbia*

[2]*Institute of Chemical Research of Catalonia, ICIQ, The Barcelona Institute of Science and Technology, Av. Països Catalans, 16, 43007 Tarragona, Spain*

[3]*Serbian Academy of Sciences and Arts, Knez Mihajlova 35, 11000 Belgrade, Serbia*

[4]*Department of Physics and Astronomy, Uppsala University, Box 516, 751 20 Uppsala, Sweden*

[5]*Department of Materials Science and Engineering, School of Industrial Engineering and Management, KTH - Royal Institute of Technology, Brinellvägen 23, 100 44 Stockholm, Sweden*

* **corresponding author**, e-mail: igor@ffh.bg.ac.rs

**Increasing demand and high prices of advanced catalysts motivate a constant search for novel active materials with reduced content of noble metals. The development of thin films and core-shell catalysts seem to be a promising strategy along this path. Using Density Functional Theory we have analyzed a number of surface properties of supported bimetallic thin films with composition $A_3B$ (where A = Pt, Pd, B = Cu, Ag, Au). We focus on surface segregation, dissolution stability and surface electronic structure. We also address the chemisorption properties of $Pd_3Au$ thin films supported by different substrates, by probing the surface reactivity with CO. We find a strong influence of the support in the case of mono- and bilayers, while the surface strain seems to be the predominant factor in determining the surface properties of supported trilayers and thicker films. In particular, we show that the studied properties of the supported trilayers can be predicted from the lattice mismatch between the overlayer and the support. Namely, if the strain dependence of the corresponding quantities for pure strained surfaces is known, the properties of strained supported trilayers can be reliably estimated. The obtained results can be used in the design of novel catalysts and predictions of the surface properties of supported ultrathin catalyst layers.**

**Keywords:** thin film catalyst; surface strain; surface segregation; catalyst stability; catalyst design

# 1. Introduction

Modern materials science and catalysis are facing challenges concerning metal utilization, technical possibilities and economic justification. The necessity of having an efficient catalyst in most cases keeps the selection of the active component focused on noble metals.[1-3] However, their high and often volatile price provides a motivation for reducing their content to the lowest possible limit and for the development of new approaches for rational catalyst design.[4-6] For this reason, there is a high interest in thin film and even single atom catalysts.[7-10]

The activity of a heterogeneous catalyst is dictated mainly by the surface. This implies that the underlying bulk material is much less important, unless it significantly affects the chemical properties of the surface layer. Hence, a number of thin film (TFC) and core-shell catalysts (CSC) have been developed. It has been governed by the idea of substituting the bulk of an expensive noble catalyst with a cheaper material that does not interfere with the surface properties. Both TFC and SCS are composed of a coating or shell grown over a support or core. Coating/shell bears the catalytic function and can consist of a single catalytically active metal or a multi-metal phase, which can further boost the catalytic performance.[11,12] While a clear border between the bulk and surface is difficult to draw, some studies have shown that many properties of a solid are close to those of bulk already three layers below the surface.[13]

Concerning practical applications, a catalyst must have high activity and stability under operating conditions. The activity is determined by the chemical composition of the surface and its electronic structure.[14-19] In a bimetallic system additional surface processes can take place, such as surface segregation,[20-22] which affects the chemical composition and, ultimately, the catalyst activity of the surface. Considering the catalyst stability, a catalyst can undergo corrosion, dissolution or various aggregation processes, which also can lead to an activity loss.[23]

In the present work we address the properties of thin bimetallic films on different substrates, as models of TFC or CSC, focusing on compounds of Pt (Pd) and coinage metals (Cu, Ag and Au). These systems have been shown to be suitable catalysts for a number of catalytic processes. The studies of Pd-Cu alloys have demonstrated correlations between the composition of the alloy, on one hand, and its electronic and adsorption properties, on the other.[24] Moreover, due to some cooperative effects the Pd-Cu surfaces display higher reactivity than it could be expected based on the *d*-band model.[25] Pd-Au catalysts have been in focus of materials science research for years.[26-28] Similarly, Pt-based bimetallic catalysts have widely been investigated as well, showing enhanced activity.[29-34] In order to link the activity and composition of nanoparticles and polycrystalline alloys based on Pt, significant efforts have been made in organizing and analyzing the available experimental results.[35,36]

In TFC (or CSC) the surface composition and the electronic structure of the coating (shell) can be affected by underlying support (core). The effects can be due to the chemical environment (ligand effect) or due to the lattice mismatch (strain effect). Elastic deformation has long been a suitable tool for modifying surface reactivity[37] since outside the elastic regime different types of relaxation may happen.[38] The understanding of the changes of different properties of thin catalyst film with its thickness and the ability to predict such properties are of great importance for practical applications. Hence, here we analyze the segregation trends, stabilities, in terms of dissolution, and the electronic structure of the $A_3B$ (where A = Pt, Pd, and B = Cu, Ag, Au) systems supported by WC(0001) and compare them to those obtained for corresponding pure strained alloy surfaces. We have chosen WC as a support as it has a very strong ligand effect.[39,40] Also, we compare the properties of thin $Pd_3Au$ layers supported by different substrates (Pd, Ag, Pt, Au, and WC) with those of pure strained $Pd_3Au$ surface. For the case $Pd_3Au$ overlayers we investigate the reactivity, which we probe with CO, chosen due to its importance in many catalytic processes.[41] Also, we study segregation trends under the conditions of CO adsorption. We show that the effect of the support is largely lost already for trilayers and that their properties can be predicted by pure strained alloy surfaces.

## 2. Computational Details

The calculations were performed using the PWscf code as implemented in the Quantum ESPRESSO distribution[42] within the Perdew–Burke–Ernzerhof[43] (PBE) functional. The plane wave kinetic energy cut-off was 36 Ry while the charge density cut-off was 576 Ry. First irreducible Brillouin zone was sampled using a $\Gamma$-centered 4×4×1 *k*-point grid.

Selected bimetallic systems were constructed of Pt or Pd (metal A) and Cu, Ag or Au (metal B) with $L_1 2$ structure of the composition $A_3B$. The pure strained compounds were modeled as 5-layer slabs of the (111) orientation. The WC-supported thin film catalysts were modelled as bi- and trilayers epitaxially grown on six layers of WC with the (0001) orientation (Fig. 1). The overlayers contained 4 atoms *per* layer. Additionally, $Pd_3Au$ overlayers were also modelled using 3 layer thick slabs on other metallic supports such as Pd, Ag, Pt and Au. In all the calculations, the bottom two layers of the substrate were fixed, while the others were fully relaxed. The vacuum thickness was around 15 Å that together with applied dipole correction[44] minimized the interaction of periodic images along the *z*-axis of the simulation cell.

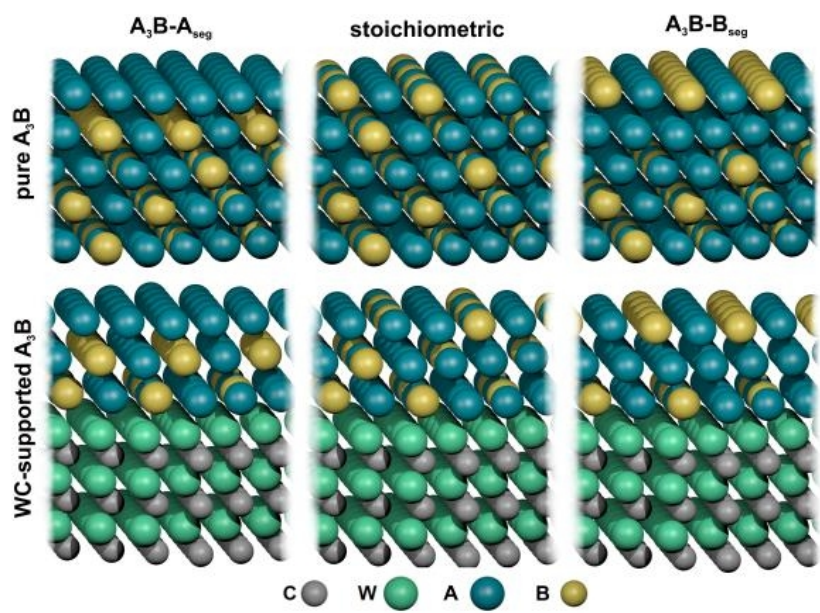

**Figure 1.** Modelled structures of pure (top row) and WC-supported (bottom row) $A_3B$ bimetallic systems undergoing surface segregation processes. $A_3B-B_{seg}$ denotes the surface where B (B = Cu, Ag or Au) segregates to the surface while $A_3B-A_{seg}$ denotes the surface where A (A = Pd or Pt) segregates to the surface. In the stoichiometric slab all layers retain the composition $A_3B$.

We have calculated several characteristics of the investigated overlayers. First, we calculated the segregation energies ($E_{seg}$)[22], using the simplified scheme suggested in Ref. [45]:

$$E_{seg} = E_{X\text{-}seg} - E_{stoich} \qquad (1)$$

where $E_{X\text{-}seg}$ and $E_{stoich}$ stand for the total energy of the segregated slab (X = A or B) and the total energy of the stoichiometric slab, respectively. Naturally, $E_{seg}$ cannot be calculated for a monolayer.

Next, in order to estimate the stability of supported layers with respect to the dissolution/corrosion processes we calculated the energy of dissolution ($E_{diss}$) which we defined as:

$$E_{diss} = E_{intact} - E_{X\text{-}dissolved} - E_{X,bulk} \qquad (2)$$

where $E_{intact}$, $E_{X\text{-}dissolved}$ and $E_{X,bulk}$ stand for the total energy of the surface layer before removal of an atom, the total energy upon removal of X (A or B) from the surface layer and the energy of atom X in its bulk phase, respectively. In physical terms, this process actually refers to the formation of the vacancy in the surface layer. However, considering catalyst stability, a parallel can be drawn to catalysts dissolution under operating conditions which is of great importance when using catalysts under wet conditions or in electrochemical processes. We chose to refer the energy of the above process with respect to the bulk phase of the metal that is removed from the surface layer (i.e. dissolved). In this case one can immediately compare the stability of a given metal in the catalyst layer and estimate whether it will show higher tendency towards dissolution as compared to pure metallic phase. This energy can be calculated for the catalyst layer of any thickness. For the studied bimetallic surfaces it was calculated only for the most stable segregated configurations. Finally, for the investigated surfaces we calculated the $d$-band centers ($E_{d\text{-}band}$), which are well accepted as reactivity descriptors.[15] Reported $E_{d\text{-}band}$ are average $E_{d\text{-}band}$ of the all the surface atoms of a given type (A or B).

In addition to the listed properties, we also investigated the chemisorption of CO on supported $Pd_3Au$ layers. The CO adsorption was tested at all high symmetry sites[46] and quantified using the CO adsorption energy ($E_{ads,CO}$), defined as:

$$E_{ads,CO} = E_{slab+CO} - E_{slab} - E_{CO} \qquad (3)$$

In the equation above $E_{slab+CO}$, $E_{slab}$ and $E_{CO}$ stand, respectively, for the total energy of a slab with adsorbed CO, the total energy of the slab and the total energy of the isolated CO molecule. Besides, we calculated the segregation energies under the conditions of CO adsorption ($E_{seg,CO}$) defined as:

$$E_{seg,CO} = E_{X-seg,CO} - E_{stoich,CO} \qquad (4)$$

where $E_{X-seg,CO}$ and $E_{stoich,CO}$ have the same meaning as for Eq. (1) but with CO adsorbed on the surface. We note that the three possible types of surfaces (stoichiometric, A- and B-segregated) have different preferential sites for the CO adsorption and we use Eq. (4) to calculate the energies of the most stable configurations.

In order to compare the calculated characteristics of the supported bimetallic layers to those of pure bimetallic systems we performed similar calculations for strained bimetallic $A_3B(111)$ surfaces, represented by 5-layer slabs. In these calculations the lateral strain was varied in the range of ±6%. We fitted the results using the fourth order polynomial function. The obtained strain dependences of the calculated properties were further applied to predict the properties for any given surface strain within the mentioned strain range.

## 3. Results and Discussion

The calculated lattice parameters of the investigated $A_3B$ systems are provided in Table S1 (Supplementary Information). The calculated lattice parameters of the $A_3B$ compounds with B=Cu are smaller than those of pure Pt and Pd and when B=Ag, Au the trend is opposite. When placed onto the WC(0001) surface all the $A_3B$ systems are under tensile strain, depending on the lattice mismatch between them and WC (Table 1). $Pd_3Ag$, $Pd_3Au$ as well as $Pt_3Ag$ and $Pt_3Au$ are strained by about 2.8%, while for the compounds with copper the strain is above 5%. For the $Pd_3Au$ overlayers on different substrates, the strain varies in a wider range, and the overlayer experiences both compressive and tensile surface strain (Table 2). The calculated segregation energies, dissolution energies and *d*-band centers for all the considered supported trilayers are summarized in Tables 1 (trilayers on WC(0001)) and 2 (Pd$_3$Au trilayers on different substrates).

**Table 1.** Calculated properties of supported A$_3$B catalysts trilayers on WC(0001) substrate. Columns define the composition, while rows give segregation ($E_{seg}$) and dissolution energy ($E_{diss}$) or the *d*-band center position ($E_{d\text{-band}}$) for A (Pt or Pd) or B (Cu, Ag or Au). $E_{diss}$ and $E_{d\text{-band}}$ are given for the segregated surfaces with negative segregation energies.

| @WC | | A = Pd | | | A = Pt | | |
|---|---|---|---|---|---|---|---|
| | | B = Ag | B = Au | B = Cu | B = Ag | B = Au | B = Cu |
| **Strain / %** | | 2.96 | 2.74 | 5.82 | 2.91 | 2.57 | 5.48 |
| $E_{seg}$ / eV | A | 0.29 | 0.36 | −0.01 | 0.20 | 0.27 | −0.09 |
| | B | −0.34 | −0.19 | 0.05 | −0.56 | −0.52 | 0.08 |
| $E_{diss}$ / eV | A | −0.53 | −0.58 | −0.80 | −0.45 | −0.42 | −0.97 |
| | B | −1.06 | −1.13 | / | −1.03 | −0.97 | / |
| $E_{d\text{-band}}$ / eV | A | −1.65 | −1.71 | −1.76 | −2.15 | −2.23 | −2.28 |
| | B | −3.31 | −2.93 | / | −3.36 | −3.05 | / |

Further, we analyse the segregation energies obtained for the pure strained A$_3$B(111) surfaces without any substrate (data for these surfaces are given in Table S2, Supplementary Information). It can be concluded that in pure bimetallic systems there is a strong tendency towards surface segregation of Ag and Au while in the case of copper compounds Pd and Pt tend to segregate on the surface. These results agree with previous reports.[21,45] It can be seen that the segregation energy depends on the strain but specifically for each A-B combination, so no general rule can be derived regarding its strain dependence. From the $E_{seg}$ *vs*. strain dependence for pure A$_3$B(111) surfaces (Table S2, Figure S1), we have predicted surface segregation energies for metallic overlayers with the strain corresponding to that occurring when the films are supported by WC(0001) and for the Pd$_3$Au overlayers on different substrates, and compared these values to the ones reported in Table 1 and Table 2.

**Table 2.** Calculated properties of supported Pd$_3$Au catalysts trilayers on different substrates defined in columns. Rows give segregation ($E_{seg}$), dissolution energy ($E_{diss}$), d-band centers ($E_{d\text{-band}}$), segregation energies in the presence of CO ($E_{seg,CO}$) and CO adsorption energies ($E_{ads,CO}$) for the studied surfaces. Rows define the element which segregates to the surface.

| Pd$_3$Au@ | | Ag(111) | Au(111) | Pd(111) | Pt(111) | WC(0001) |
|---|---|---|---|---|---|---|
| **Strain %** | | 3.41 | 3.63 | −1.24 | −0.80 | 2.74 |
| $E_{seg}$ / eV | Pd | 0.29 | 0.30 | 0.33 | 0.36 | 0.36 |
| | Au | −0.24 | −0.25 | −0.28 | −0.29 | −0.19 |
| $E_{diss}$ / eV | Pd | −0.60 | −0.59 | −0.61 | −0.61 | −0.58 |
| | Au | −1.13 | −1.13 | −1.10 | −1.12 | −1.13 |
| $E_{d\text{-band}}$ / eV | Pd | −1.60 | −1.58 | −1.83 | −1.79 | −1.71 |
| | Au | −2.88 | −2.87 | −3.13 | −3.11 | −2.93 |
| $E_{seg,CO}$ / eV | Pd | 0.37 | 0.38 | 0.36 | 0.38 | 0.32 |
| | Au | 0.19 | 0.20 | 0.13 | 0.15 | 0.22 |
| $E_{ads,CO}$ / eV | Pd | −2.10 | −2.10 | −1.93 | −1.97 | −2.20 |
| | stoich | −2.18 | −2.18 | −1.96 | −1.99 | −2.15 |
| | Au | −1.75 | −1.73 | −1.55 | −1.55 | −1.74 |

The predicted values are also compared with those explicitly calculated for the supported bilayers. The comparison for all the supported overlayers is given in Fig. 2. As can be seen, there is good agreement between the "predicted" and calculated values of the segregation energies for the supported trilayers, which is, generally, not the case for the bilayers. Although there are certain deviations of the "predicted" values from the calculated ones, the prediction of the sign of the segregation energy, in other word the element which segregates at the surface, is always in agreement with the explicitly calculated segregation energy. Hence, upon reaching the thickness of three layers in the coating/shell of TFC/CSC one can predict surface composition on the basis of the behavior of pure strained bimetallic compounds that constitute the coating/shell.

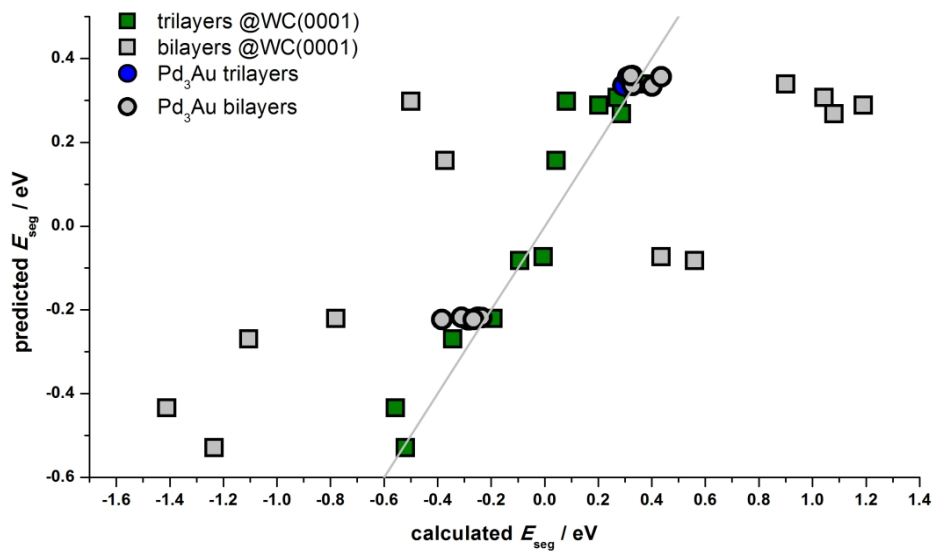

**Figure 2.** Correlation of the calculated surface segregation energies for the supported bi- and trilayers and the segregation energies predicted on the basis of the strain dependence of $E_{seg}$ of pure $A_3B(111)$ surfaces (Table S2). Gray line represents the $E_{seg}$(predicted) = $E_{seg}$(calculated) depende ($R^2$(trilayers) = 0.94).

Next, we turn to the dissolution stability issues of supported overlayers. The dissolution of surface atoms was modelled as a vacancy formation by removing one atom of a particular kind (A or B) from the surface. The most stable segregated surfaces were tested for dissolution stability (see Tables 1 and 2 for the results for supported trilayers). Similar calculations were done for segregated strained $A_3B(111)$ surfaces modelled as 5-layer slabs (Table S3, Supplementary Information). We see that all the dissolution energies are negative that, according to Eq. (2), indicates that the studied bimetallic systems are less prone to dissolution than the corresponding pure metallic phases. Moreover, there is a very pronounced dependence of $E_{diss}$ on the lateral strain. As surface gets more compressed, $E_{diss}$ gets more positive indicating a destabilization of the lattice. Using the $E_{diss}$ vs. strain dependencies for pure $A_3B(111)$ surfaces (Table S3, Figure S2, Supplementary Information) we predicted the dissolution energies for supported overlayers. We again compared the calculated and predicated $E_{diss}$ (Fig. 3) and obtained exceptionally good agreement in the case of supported trilayers. In the case of supported bilayers the agreement between the calculated and predicted values is worse.

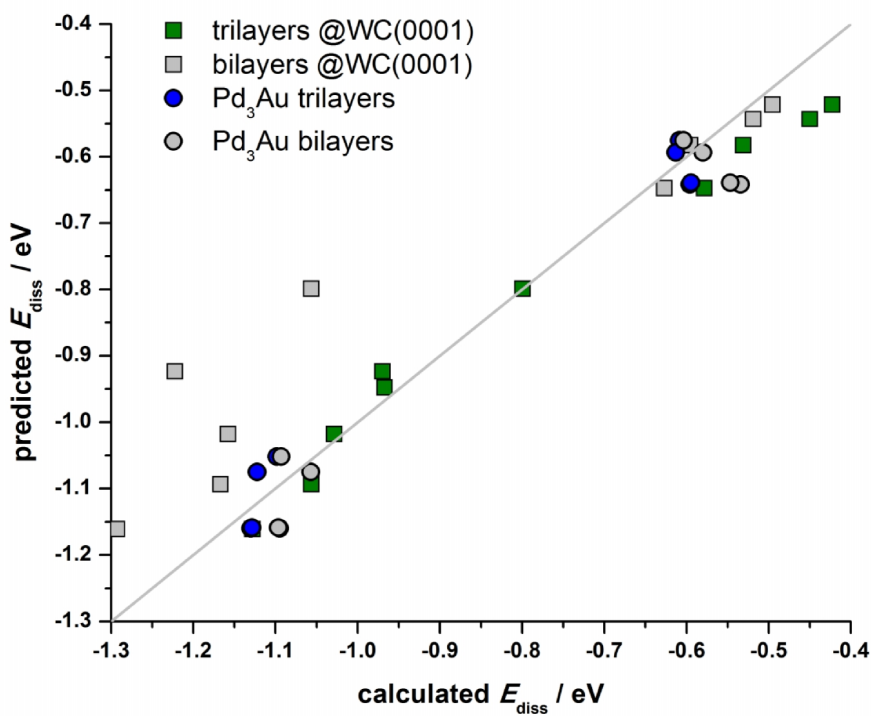

**Figure 3.** Calculated dissolution energies for the supported bi- and trilayers correlated to the dissolution energies predicted on the basis of the strain dependence for the case of $E_{diss}$ of pure $A_3B(111)$ surfaces. Gray line represents the $E_{diss}$(predicted) = $E_{diss}$(calculated) dependence ($R^2$(trilayers) = 0.96).

Finally, we calculated the *d*-band centers of the supported trilayers (Tables 1 for trilayers on WC(0001) and Table 2 for Pd$_3$Au trilayers on different substrates). This parameter is well accepted as an activity descriptor and is known to strongly depend on the surface strain.[15,37] Calculated $E_{d-band}$ for the most stable segregated $A_3B(111)$ surfaces are given in Table S4 (Supplementary Information). The well-known behavior is clearly seen: as lattice gets more compressed the *d*-band centers shift to lower values[15,47] Moreover, a clear distinction between the values of the *d*-band centers of A (Pd and Pt) and B (Cu, Ag, Au) is visible, the latter ones having *d*-bands filled and located far from the Fermi level. Again, based on the strain dependence of $E_{d-band}$ for pure $A_3B(111)$ (Table S4, Figure S3) we estimated the values of $E_{d\text{-band}}$ for the strained overlayers on WC(0001) and Pd$_3$Au overlayers, using the exact values of strain these bimetallic systems experience on the corresponding supports. The comparison between the predicted values and explicitly calculated $E_{d\text{-band}}$ for the bi- and trilayers including all the studied surfaces, irrespective of preferred segregation, are shown in Fig. 4. We observe an extremely good correlation between the calculated and predicted values, even for the case of bilayers.

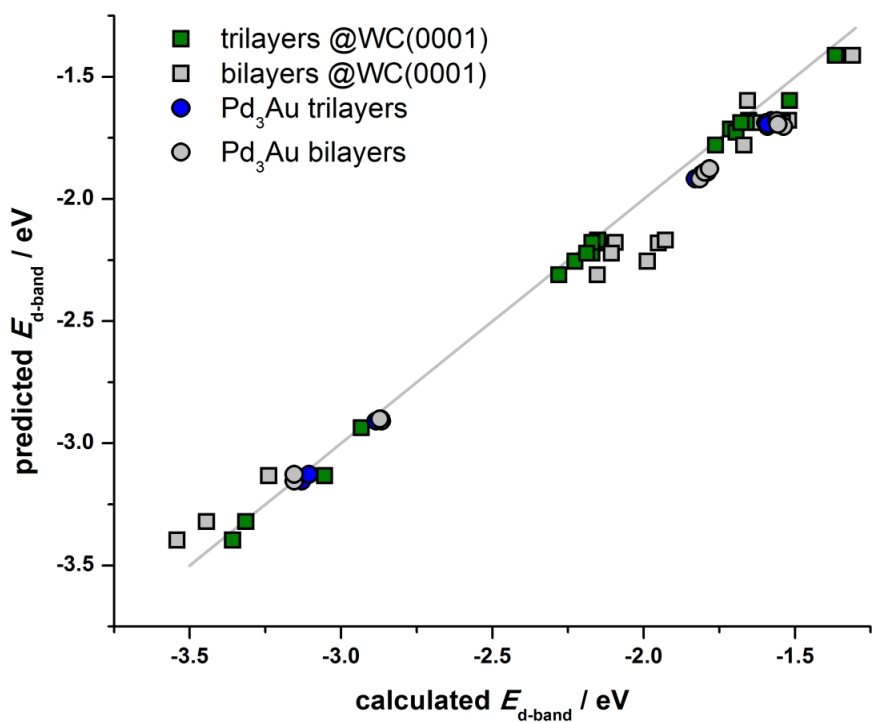

**Figure 4.** Calculated *d*-band centers for the supported bi- and trilayers correlated to the *d*-band centers predicted on the basis of the strain dependence of $E_{d\text{-band}}$ of pure A$_3$B(111) surfaces. Gray line represents the $E_{d\text{-band}}$(predicted) = $E_{d\text{-band}}$(calculated) dependence ($R^2$(trilayers) = 0.998).

Based on the results presented so far it can be concluded that the chemical composition, dissolution stability and electronic structure of the supported layers can be estimated on the basis of lattice mismatch between the overlayer and the substrate (or surface strain) already for trilayers, using pure strained surfaces as models. For thinner overlayers the ligand effect of the support is very pronounced and it significantly affects the surface properties. Although not presented here, we evaluated the listed properties (where applicable) for monolayers as well. In this case, no correlation with predicted values was found due to a very strong influence of the support and, therefore, an explicit modelling should be performed for each system in question.

In addition, we investigated whether a similar prediction could be used for the reactivity and segregation behaviour in a selected atmosphere. We performed such an analysis for the case of the supported $Pd_3Au$ trilayers using CO as a probe. CO adsorption was investigated on segregated (both Pd and Au) and stoichimetric $Pd_3Au$ overlayers on 5 different supports (15 surfaces in total). We considered all possible adsorption sites and identified the preferred ones. As indicated by the previous work,[48] three-fold sites were most favored. The reactivity of these surfaces is greatly influenced by the chemical composition and also surface strain. $E_{ads,CO}$ was found to be between −2.15 and −1.55 eV. We find that in the presence of CO the surface segregation of both Pd and Au is not favored in all the cases and that the surfaces with stoichiometric composition are by 0.20 eV to 0.35 eV lower in energy than the segregated surfaces. Further, we calculated the CO adsorption energies for pure $Pd_3Au(111)$ surface (Pd-segregated, Au-segregated and stoichiometric one) and also the segregation energies under the conditions of CO adsorption as a function of strain (Table S5, Supplementary Information). Finally, these dependencies were used to predict $E_{ads,CO}$ and $E_{seg,CO}$ for strained $Pd_3Au$ overlayers based on the amount of strain they experience on studied substrates. The obtained correlations between the predicted and explicitly calculated $E_{ads,CO}$ and $E_{seg,CO}$ for supported $Pd_3Au$ trilayers are given in Fig. 5. Again we see that the properties of supported trllayers are well described using strained pure surfaces. Relative errors of predicted $E_{ads,CO}$ are up to 5% of the calculated values.

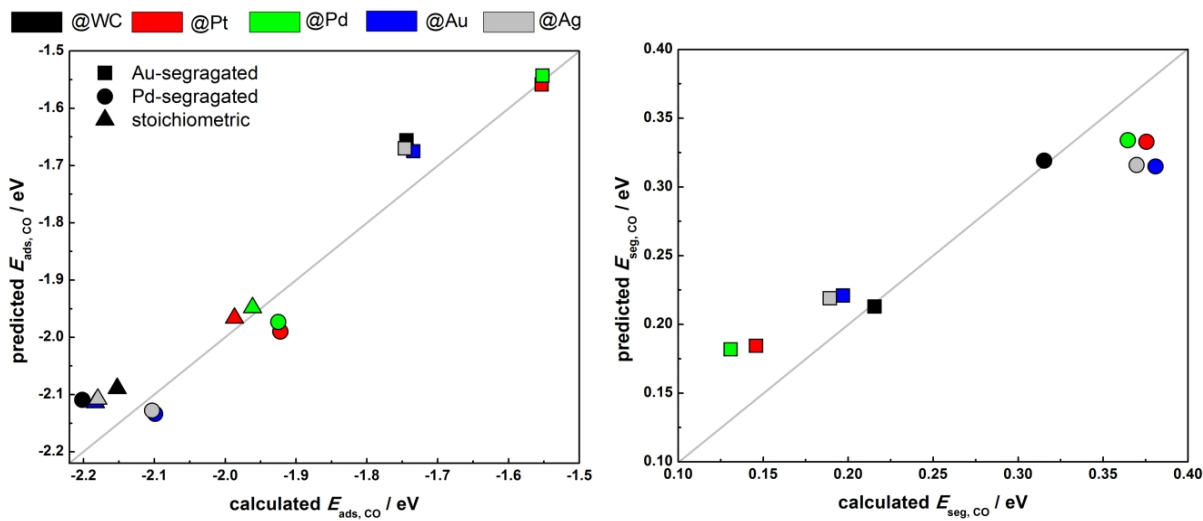

**Figure 5.** Comparison of the calculated and predicted properties for $Pd_3Au$ trilayers on different substrates: CO adsorption energies (left) and surface segregation energies in the presence of CO (right). Surface composition is defined by the symbol, while color defines the substrate (coefficients of determination $R^2(E_{ads,CO})$ = 0.99; $R^2(E_{seg,CO}$ = 0.94)).

Finally, while the present calculations are performed for the densely packed (111) surfaces of FCC metals/compounds, we cannot foresee any physical limitations to apply similar ideas to other surfaces as well. While the application of the presented ideas for thin film catalysts is more or less straightforward, the situation with the nanosized core-shell systems might be more complicated. In the case of very small clusters and nanoparticles the local coordination of atoms and chemical environment can play a dominant role in determining stability and catalytic activity. However, in the case of larger nanoparticles where a significant fraction of well-defined surface planes can be identified, we expect that the strain of the shell layer can be used to predict the shell properties. At the same time, the possibility of an edge formation in order to relieve the strain must not be disregarded.[49] However, the results regarding line defects in bimetallic Pt-Cu nanoparticles suggest that the effect of line defects (twin boundary, in particular case) is lost already after several atomic layers.[50] In this sense, it might be anticipated that nanoparticle surface facets behave like extended surfaces when far enough from the edge/defect.

## 4. Conclusions

We have demonstrated that a number of surface properties of thin supported films can be reliably predicted using strained pure surfaces as models, considering that the thickness of supported film is sufficiently large and the distortion is within few %. The investigated properties include segregation energies, dissolution energies and $d$-band centers. In specific, for the bimetallic overlayers with thickness of 3 atomic layers we find that the segregation energies and dissolution energies are well estimated using the strain dependence of the same properties for the corresponding pure bimetallic surfaces of the same orientation. In addition, the predicted values of the $d$-band centers also show exceptionally good agreement with the calculated ones. For $Pd_3Au$ trilayers on different substrates we show that the same conclusions hold for the CO adsorption energies and segregation energies in the presence of CO. The obtained results offer a simple strategy for *a priori* predictions of surface properties of supported few-atoms thick catalyst layers and can be used for rational design of novel thin film or core-shell catalysts. Further generalization and formation of extensive databases containing surface properties of various catalysts is necessary to exploit the full potential of the presented findings.

**Conflicts of interest**

There are no conflicts to declare.

**Acknowledgement**

This work was supported by the Swedish Research Links initiative of the Swedish Research Council (348-2012-6196). I.A.P. and A.S.D. acknowledge the support provided by the Serbian Ministry of Education, Science and Technological Development through the project III45014. N.V.S. acknowledges the support provided by Swedish Research Council through the project No. 2014-5993. Financial support provided through the NATO Project EAP.SFPP 984925 - "DURAPEM - Novel Materials for Durable Proton Exchange Membrane Fuel Cells" is also acknowledged. We also acknowledge the support from Carl Tryggers Foundation for Scientific Research. The computations were performed on resources provided by the Swedish National Infrastructure for Computing (SNIC) at National Supercomputer Centre (NSC) at Linköping University and High Performance Computing Center North (HPC2N) at Umeå University. We thank MINECO (CTQ2015-68770-R) for financial support. E. F. thanks MINECO La Caixa – Severo Ochoa for a predoctoral grant through Severo Ochoa Excellence Accreditation 2014–2018 (SEV-2013-0319).

**SUPPLEMENTARY INFORMATION**

1. Lattice constants of investigated bimetallic systems

**Table S1.** Calculated lattice parameters of investigated $A_3B$ bimetallic systems

| $A_3B$ | $a_0$ / Å |
|---|---|
| $Pt_3Cu$ | 3.923 |
| $Pt_3Ag$ | 4.020 |
| $Pt_3Au$ | 4.034 |
| $Pd_3Cu$ | 3.910 |
| $Pd_3Ag$ | 4.019 |
| $Pd_3Au$ | 4.027 |

## 2. Strain dependence of the segregation energies for the case of pure A$_3$B(111) surfaces

**Table S2.** Dependence of surface segregation (in eV) energies of A (Pt, Pd) and B (Cu, Ag, Au) on the lateral surface strain in A$_3$B(111) surfaces. At the end parameter of the polynomial fit of $E_{seg}$ vs. strain (S, in %) are given.

| | segregates | strain / % | | | | |
|---|---|---|---|---|---|---|
| | | 6 | 3 | 0 | −3 | −6 |
| Pd$_3$Cu | A | −0.07 | −0.07 | −0.06 | −0.04 | −0.13 |
| | B | 0.15 | 0.20 | 0.25 | 0.29 | 0.34 |
| Pd$_3$Ag | A | 0.30 | 0.27 | 0.22 | 0.18 | 0.17 |
| | B | −0.30 | −0.27 | −0.24 | −0.23 | −0.22 |
| Pd$_3$Au | A | 0.32 | 0.34 | 0.35 | 0.37 | 0.41 |
| | B | −0.21 | −0.22 | −0.22 | −0.22 | −0.22 |
| Pt$_3$Cu | A | −0.07 | −0.13 | −0.18 | −0.23 | −0.28 |
| | B | 0.28 | 0.40 | 0.50 | 0.55 | 0.57 |
| Pt$_3$Ag | A | 0.40 | 0.29 | 0.14 | 0.01 | −0.07 |
| | B | −0.51 | −0.44 | −0.39 | −0.39 | −0.48 |
| Pt$_3$Au | A | 0.35 | 0.31 | 0.26 | 0.21 | 0.22 |
| | B | −0.56 | −0.53 | −0.51 | −0.50 | −0.56 |
| | segregates | fit* | | | | |
| | | A0 | B1 | B2 | B3 | B4 |
| Pd$_3$Cu | A | −5.86E−02 | −7.85E−03 | 1.20E−03 | 3.38E−04 | −6.59E−05 |
| | B | 2.47E−01 | −1.57E−02 | −4.56E−05 | 4.35E−07 | 2.07E−06 |
| Pd$_3$Ag | A | 2.25E−01 | 1.59E−02 | −2.78E−05 | −1.47E−04 | 1.14E−05 |
| | B | −2.44E−01 | −6.64E−03 | −5.95E−04 | −3.16E−06 | 3.49E−06 |
| Pd$_3$Au | A | 3.52E−01 | −4.96E−03 | 2.75E−04 | −8.20E−05 | 3.63E−06 |
| | B | −2.23E−01 | 2.20E−04 | 3.33E−04 | 1.83E−05 | −3.52E−06 |
| Pt$_3$Cu | A | −1.81E−01 | 1.62E−02 | 1.53E−04 | 4.00E−05 | −1.10E−06 |
| | B | 5.00E−01 | −2.73E−02 | −2.96E−03 | 8.30E−05 | 2.60E−05 |
| Pt$_3$Ag | A | 1.43E−01 | 5.08E−02 | 5.61E−04 | −3.24E−04 | 4.59E−06 |
| | B | −3.92E−01 | −1.01E−02 | −1.72E−03 | 2.08E−04 | −3.41E−05 |
| Pt$_3$Au | A | 2.58E−01 | 1.95E−02 | 4.28E−04 | −2.30E−04 | 1.08E−05 |
| | B | −5.07E−01 | −6.76E−03 | −1.22E−03 | 1.99E−04 | −5.98E−06 |

*$E_{seg}$ = A0 + B1×S + B2×S$^2$ + B3×S$^3$ + B4×S$^4$

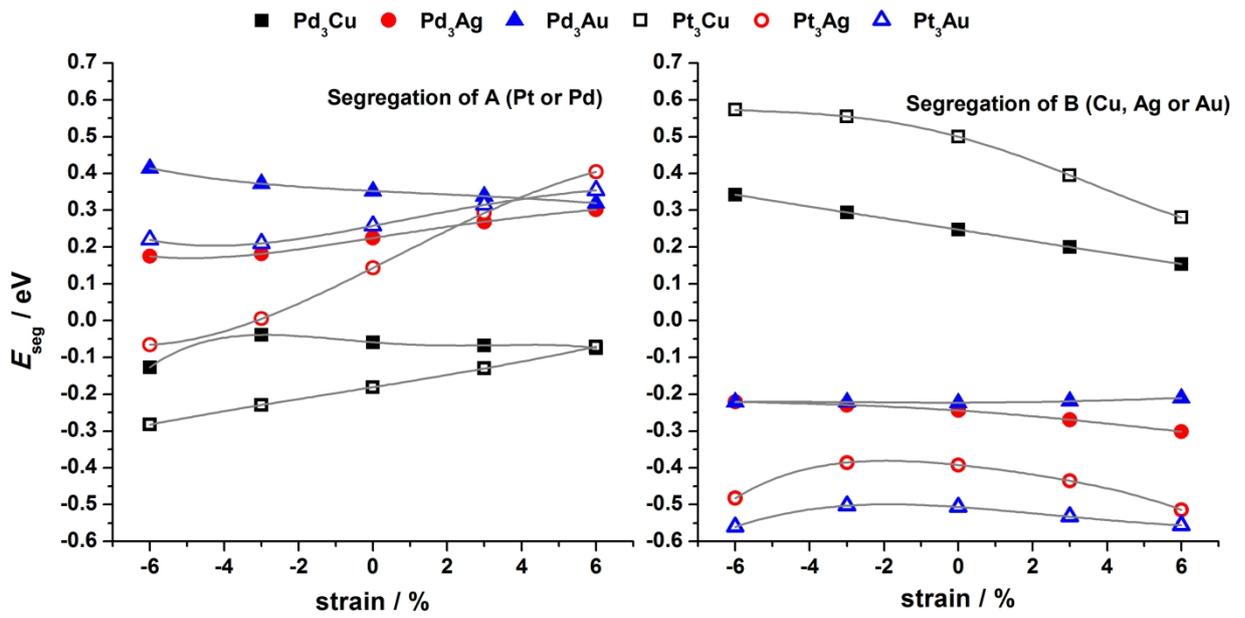

**Figure S1.** Strain dependence of A (Pt or Pd; left) and B (Cu, Ag or Au; right) segregation energies.

## 3. Strain dependence of the dissolution energies for the case of pure segregated A$_3$B(111) surfaces

**Table S3.** Dependence of dissolution energies (in eV) of given atoms (defined by rows) on the lateral surface strain in segregated A$_3$B(111) surfaces. At the end parameter of the polynomial fit of $E_{diss}$ vs. strain (S, in %) are given.

| | dissolves | strain / % | | | | |
|---|---|---|---|---|---|---|
| | | 6 | 3 | 0 | −3 | −6 |
| Pd$_3$Cu (Pd segregated) | Pd | −0.79 | −0.87 | −0.87 | −0.73 | −0.39 |
| Pd$_3$Ag (Ag segregated) | Ag | −1.07 | −1.09 | −1.03 | −0.85 | −0.52 |
| | Pd | −0.49 | −0.58 | −0.61 | −0.56 | −0.39 |
| Pd$_3$Au (Au segregated) | Au | −1.12 | −1.16 | −1.11 | −0.92 | −0.55 |
| | Pd | −0.59 | −0.65 | −0.62 | −0.47 | −0.17 |
| Pt$_3$Cu (Pt segregated) | Pt | −0.91 | −1.09 | −1.14 | −1.02 | −0.64 |
| Pt$_3$Ag (Ag segregated) | Ag | −0.97 | −1.02 | −0.94 | −0.74 | −0.46 |
| | Pt | −0.41 | −0.54 | −0.59 | −0.55 | −0.15 |
| Pt$_3$Au (Au segregated) | Au | −0.87 | −0.93 | −0.86 | −0.67 | −0.35 |
| | Pt | −0.47 | −0.52 | −0.48 | −0.30 | −0.03 |

| | dissolves | fit* | | | | |
|---|---|---|---|---|---|---|
| | | A0 | B1 | B2 | B3 | B4 |
| Pd$_3$Cu (Pd segregated) | Pd | −8.67E−01 | −2.02E−02 | 7.62E−03 | −3.64E−04 | −4.56E−07 |
| Pd$_3$Ag (Ag segregated) | Ag | −1.03E+00 | −3.99E−02 | 6.41E−03 | −1.64E−04 | −7.67E−07 |
| | Pd | −6.13E−01 | −2.86E−03 | 4.96E−03 | −1.56E−04 | −1.96E−06 |
| Pd$_3$Au (Au segregated) | Au | −1.11E+00 | −3.72E−02 | 7.48E−03 | −2.70E−04 | 3.63E−06 |
| | Pd | −6.20E−01 | −2.71E−02 | 6.82E−03 | −2.22E−04 | −1.95E−06 |
| Pt$_3$Cu (Pt segregated) | Pt | −1.14E+00 | −6.52E−03 | 9.06E−03 | −4.37E−04 | 3.00E−05 |
| Pt$_3$Ag (Ag segregated) | Ag | −9.38E−01 | −4.85E−02 | 6.92E−03 | 1.50E−04 | −1.99E−05 |
| | Pt | −5.92E−01 | 9.77E−03 | 3.89E−03 | −8.77E−04 | 1.31E−04 |
| Pt$_3$Au (Au segregated) | Au | −8.58E−01 | −4.34E−02 | 6.87E−03 | −2.85E−06 | 2.20E−06 |
| | Pt | −4.83E−01 | −3.56E−02 | 8.46E−03 | −3.38E−05 | −5.40E−05 |

*$E_{diss}$ = A0 + B1×S + B2×S$^2$ + B3×S$^3$ + B4×S$^4$

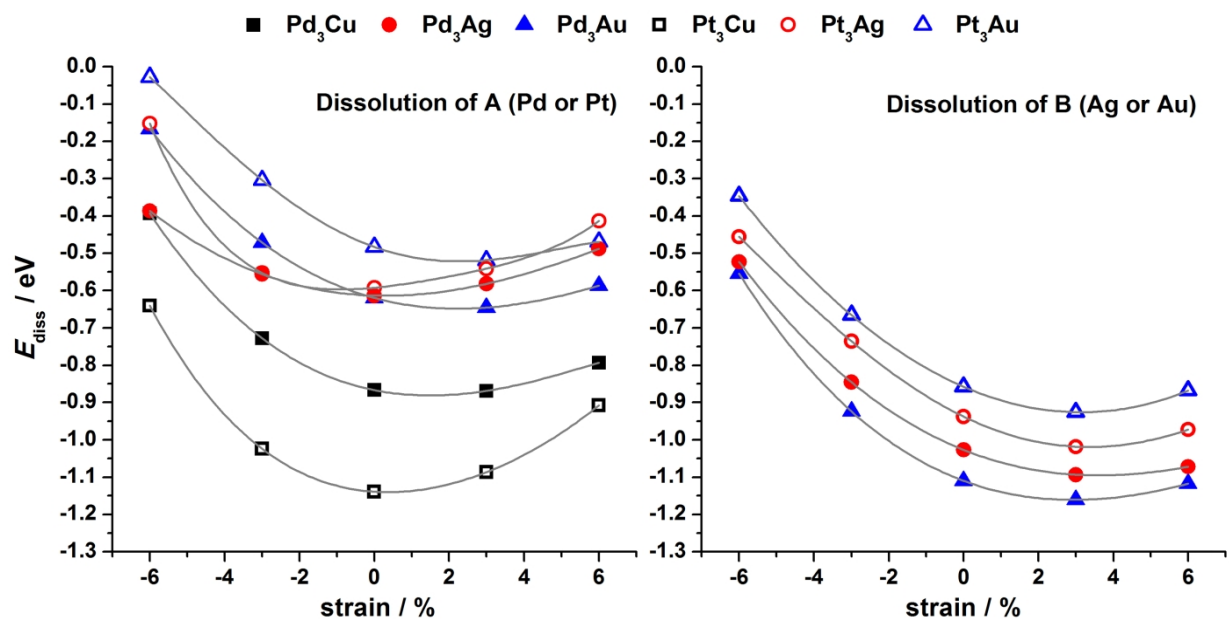

**Figure S2.** Strain dependence of $E_{diss}$ of A (Pt or Pd; left) and B (Ag or Au; right) in the most stable segregated surfaces.

## 4. Strain dependence of the d-band centers for the case of pure segregated A$_3$B(111) surfaces

**Table S4.** Dependence of the *d*-band centers (in eV) of given atoms (defined by rows) on the lateral surface strain in segregated A$_3$B(111) surfaces. At the end parameter of the polynomial fit of $E_{\text{d-band}}$ vs. strain (S, in %) are given.

|  | *d*-band | strain / % | | | | |
|---|---|---|---|---|---|---|
|  |  | 6 | 3 | 0 | −3 | −6 |
| Pd$_3$Cu (Pd segregated) | Pd | −1.77 | −1.90 | −2.05 | −2.23 | −2.32 |
| Pd$_3$Ag (Ag segregated) | Ag | −3.25 | −3.32 | −3.43 | −3.57 | −3.69 |
|  | Pd | −1.58 | −1.68 | −1.80 | −1.95 | −2.15 |
| Pd$_3$Au (Au segregated) | Au | −2.83 | −2.93 | −3.08 | −3.27 | −3.45 |
|  | Pd | −1.61 | −1.70 | −1.85 | −2.03 | −2.27 |
| Pt$_3$Cu (Pt segregated) | Pt | −2.28 | −2.45 | −2.65 | −2.89 | −3.13 |
| Pt$_3$Ag (Ag segregated) | Ag | −3.35 | −3.39 | −3.48 | −3.57 | −3.55 |
|  | Pt | −2.07 | −2.18 | −2.32 | −2.49 | −2.62 |
| Pt$_3$Au (Au segregated) | Au | −3.05 | −3.12 | −3.22 | −3.34 | −3.32 |
|  | Pt | −2.10 | −2.23 | −2.39 | −2.59 | −2.77 |
|  | *d*-band | fit* | | | | |
|  |  | A0 | B1 | B2 | B3 | B4 |
| Pd$_3$Cu (Pd segregated) | Pd | −1.93E+00 | 4.32E−02 | 2.94E−03 | −2.78E−05 | −8.72E−05 |
| Pd$_3$Ag (Ag segregated) | Ag | −3.43E+00 | 4.31E−02 | −1.73E−03 | −1.76E−04 | 1.77E−05 |
|  | Pd | −1.80E+00 | 4.43E−02 | −1.63E−03 | 7.88E−05 | −8.08E−06 |
| Pd$_3$Au (Au segregated) | Au | −3.08E+00 | 5.91E−02 | −2.31E−03 | −1.96E−04 | 1.71E−05 |
|  | Pd | −1.85E+00 | 5.47E−02 | −2.37E−03 | 5.55E−06 | −7.89E−06 |
| Pt$_3$Cu (Pt segregated) | Pt | −2.65E+00 | 7.42E−02 | −2.37E−03 | −1.03E−04 | 1.84E−05 |
| Pt$_3$Ag (Ag segregated) | Ag | −3.48E+00 | 3.35E−02 | −7.71E−04 | −4.57E−04 | 4.16E−05 |
|  | Pt | −2.32E+00 | 5.44E−02 | −1.59E−03 | −2.24E−04 | 2.57E−05 |
| Pt$_3$Au (Au segregated) | Au | −3.22E+00 | 4.02E−02 | −1.45E−03 | −4.86E−04 | 6.50E−05 |
|  | Pt | −2.39E+00 | 6.05E−02 | −2.75E−03 | −1.22E−04 | 4.27E−05 |

*$E_{\text{d-band}}$ = A0 + B1×S + B2×S$^2$ + B3×S$^3$ + B4×S$^4$

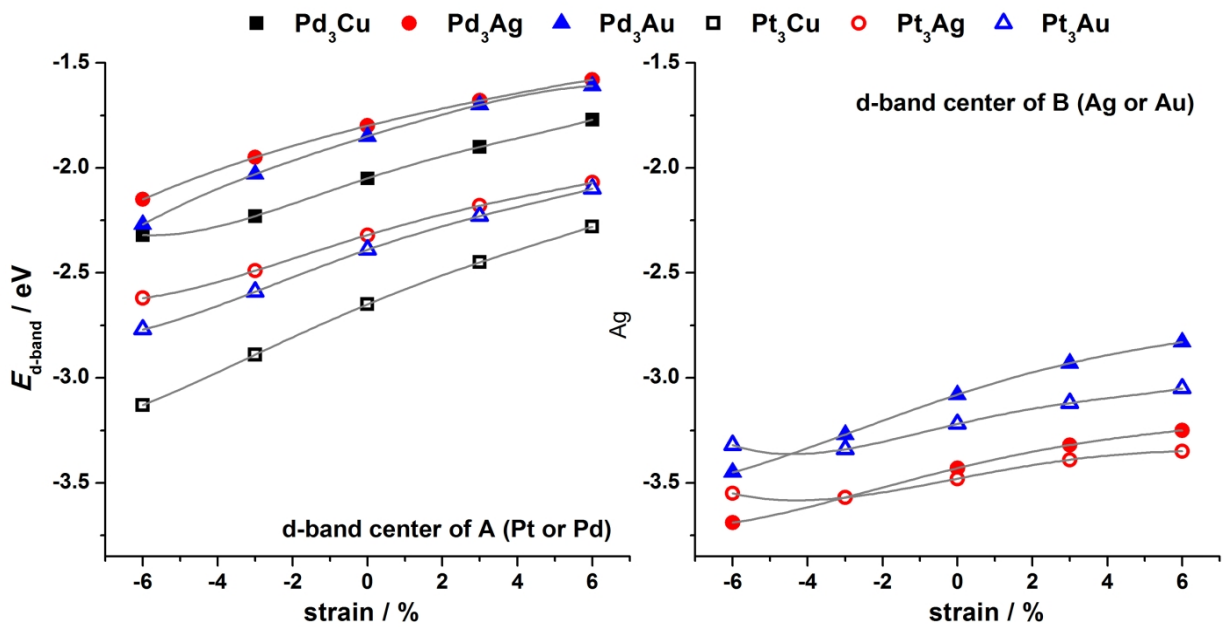

**Figure S3.** Strain dependence of $E_{\text{d-band}}$ of A (Pt or Pd; left) and B (Ag or Au; right) in the most stable segregated surfaces.

## 5. Strain dependence of CO adsorption energies and surface segregation energies in the presence of CO for pure strained A₃B(111) surfaces

**Table S5.** Dependence of the CO adsorption energies ($E_{ads,CO}$, in eV) and the surface segregation energies in the presence of CO ($E_{seg,CO}$, in eV) on the lateral surface strain in Pd$_3$Au(111) surfaces.

|  |  | strain / % | | | | |
|---|---|---|---|---|---|---|
|  | surface termination | 6 | 3 | 0 | −3 | −6 |
| $E_{ads,CO}$ / eV | Pd-segregated | −2.18 | −2.12 | −2.02 | −1.90 | −1.75 |
|  | stoichiometric | −2.17 | −2.10 | −2.00 | −1.86 | −1.67 |
|  | Au-segregated | −1.73 | −1.66 | −1.58 | −1.47 | −1.32 |
|  | segregates |  |  |  |  |  |
| $E_{seg,CO}$ / eV | Au | 0.23 | 0.22 | 0.19 | 0.17 | 0.13 |
|  | Pt | 0.31 | 0.32 | 0.33 | 0.34 | 0.33 |